\documentclass[fleqn,10pt]{wlscirep}
\usepackage[utf8]{inputenc}
\usepackage{amsmath}
\usepackage{verbatim}
\usepackage{cite}
\usepackage[dvipsnames]{xcolor}
\usepackage[T1]{fontenc}
\usepackage{tikz}
\usepackage{algorithm}%
\usepackage{algorithmicx}%
\usepackage{algpseudocode}%
\usepackage{listings}%

\title{Accumulated Local Effects and Graph Neural Networks for link prediction}

\author[1,2*]{Paulina Kaczyńska}
\author[3,4]{Julian Sienkiewicz}
\author[1]{Dominik Śl\c{e}zak}
\affil[1]{University of Warsaw, Faculty of Mathematics, Informatics and Mechanics, Institute of Informatics, Banacha 2, 02-097 Warsaw, Poland}
\affil[2]{Polish Academy of Sciences, Institute of Fundamental Technological Research, Pawińskiego 5B, 02-106, Warsaw, Poland}
\affil[3]{Warsaw University of Technology, Faculty of Physics, Koszykowa 75, 00-662 Warsaw, Poland}
\affil[4]{Warsaw University of Technology, Centre for Credible AI, Rektorska 4, 00-614 Warsaw, Poland}
\affil[*]{pm.kaczynska@student.uw.edu.pl}

\begin{abstract}
We investigate how Accumulated Local Effects (ALE), a model-agnostic explanation method, can be adapted to visualize the influence of node feature values in link prediction tasks using Graph Neural Networks (GNNs), specifically Graph Convolutional Networks and Graph Attention Networks. 
A key challenge addressed in this work is the complex interactions of nodes during message passing within GNN layers, complicating the direct application of ALE. Since a straightforward solution of modifying only one node at once substantially increases computation time, we propose an approximate method that mitigates this challenge. Our findings reveal that although the approximate method offers computational efficiency, the exact method yields more stable explanations, particularly when smaller data subsets are used. However, the explanations produced with the approximate method are not significantly different from the ones obtained with the exact method. Additionally, we analyze how varying parameters affect the accuracy of ALE estimation for both approaches.
\end{abstract}
\begin{document}

\flushbottom
\maketitle
%
%
\section*{Introduction}

Modern analysis of complex networks is a relatively new interdisciplinary area that emerged at the turn of the century with the works of Watts and Strogatz \cite{Watts1998}, Barabasi and Albert \cite{Barabsi1999}, Jeong et al. \cite{Jeong2000} and Newmann \cite{Newman2001}. The common denominator of these seminal papers is handling diverse data, such as actor \cite{Watts1998}, WWW\cite{Barabsi1999}, metabolic\cite{Jeong2000}, or scientific collaboration \cite{Newman2001} networks with sizes spanning from several hundreds to a few million nodes and showing that they share similar properties, e.g., scale-free degree distribution or small-world behavior. However, the connections among the nodes are not placed at random, which instantly raises a pertinent question: can we predict links in such a network? In a more precise manner, we might seek to determine if the network's intrinsic features, i.e., the properties inherent to the network itself, provide an answer to this problem \cite{Liben-Nowell2003}. The fundamental approach of the network science -- the BA model \cite{Barabsi1999}, which is based on the preferential attachment rule, is a simple example of a local similarity measure that is proportional to the product of nodes' degrees, therefore promoting links between highly connected nodes. Other commonly used local and global similarity measures include the number of common neighbors, Jaccard's coefficient, Adamic-Adar measure, Katz index, SimRank and many others (for a comprehensive review see, e.g., Kumar et al.\cite{Kumar2020} or Arrar et al.\cite{Arrar2024}).    

The last decade has brought a paradigm shift, connected to the rapid development of embedding methods: in 2016, Grover and Leskovec\cite{Grover2016} proposed a method for creating node embeddings using a concept similar to the famous word2vec approach\cite{Mikulov2013} (that relies on expressing a word through its semantic neighborhood transformed into a low-dimensional space). The node2vec together with DeepWalk\cite{Perozzi2014} which can be seen as a specific case of node2vec\cite{Grover2016}), encompasses the information about node neighborhood, obtained through, e.g., a random walk traversing the network, and encodes it as a vector. Owing to that, both node2vec and DeepWalk significantly outperform such algorithms as common neighbors, Adamic-Adar, Jaccard's coefficient, or preferential attachment in the task of link prediction\cite{Grover2016}.

In this context, Graph Neural Networks (GNNs)\cite{Zhou_Cui_Hu_Zhang_Yang_Liu_Wang_Li_Sun_2020} can be seen as an approach that reaches one step further: they operate in a natural way on graph-structured data and learn node embeddings by iteratively aggregating information from a node’s local neighborhood. GNNs aim at extracting complex features from the graph structure, offering advantages in link prediction\cite{Zhang2018} due to their ability to capture the information of the graph structure.

However, unlike in the case of plain similarity measures, the black-box nature of GNNs (and other deep learning models) presents significant challenges to interpretability\cite{Holzinger2022,Antamis2024} (in the sense of the ``ability to provide explanations in understandable terms to a human''\cite{Zhang2021}). A wide range of post hoc explanation methods has been developed \cite{Ying_Bourgeois_You_Zitnik_Leskovec_2019,Luo_Cheng_Xu_Yu_Zong_Chen_Zhang_2020,Kosan_Huang_Medya_Ranu_Singh_2022}, many of which aim to identify subgraphs or subsets of features most relevant to a particular prediction. Notable examples include GNNExplainer \cite{Ying_Bourgeois_You_Zitnik_Leskovec_2019} and PGEExplainer \cite{Luo_Cheng_Xu_Yu_Zong_Chen_Zhang_2020}. These methods seek to uncover the nodes, edges, or features that most significantly influence the model’s decisions, thereby offering insights into the structural aspects of the graph driving a prediction. Fewer methods, however, are designed to answer questions about how changes in a node's feature value would affect a GNN's prediction. Au et al. \cite{Au_Herbinger_Stachl_Bischl_Casalicchio_2022} categorize explainability methods into two groups: feature importance and feature effect methods. Feature importance methods quantify the relevance of each feature to a model’s prediction. Well-known examples include SHAP (SHapley Additive exPlanations) \cite{Lundberg_Lee_2017}, LIME (Local Interpretable Model-agnostic Explanations) \cite{Ribeiro_Singh_Guestrin_2016}, and permutation importance. In contrast, feature effect methods aim to illustrate how variations in a feature’s value influence the model's output. Prominent examples include Accumulated Local Effects (ALE) \cite{Apley_Zhu_2020}, Partial Dependence Plots (PDPs) \cite{Friedman_2001}, and Individual Conditional Expectation (ICE) plots \cite{Goldstein_Kapelner_Bleich_Pitkin_2015}. These methods produce visualizations that represent feature effects and average predictions, explaining not just which features matter, but how they matter. Under this classification, most GNN explainability methods fall into the feature importance category.

This study explores the application of Accumulated Local Effects (ALE) \cite{Apley_Zhu_2020} to GNNs trained for link prediction tasks. ALE visualizes the impact of a specific feature’s value on the model’s output. Below are examples of questions ALE can help answer in the context of GNNs: (1) Does increasing this node’s feature value raise or lower the probability of it being classified into a particular class? (2) Would the likelihood of a link forming between two nodes increase or decrease if their feature values were different? For instance, in a citation network model trained to predict citations, ALE can reveal how an author's affiliation affects the likelihood of their work being cited. By focusing on the individual effects of features, ALE offers a valuable complement to existing GNN explainability methods, particularly due to its applicability beyond the GNN domain. We argue that such insights are important, as they provide information that complements that of feature importance methods. While the latter identify which features matter most, ALE shows how changes in feature values influence the model’s predictions.

The ALE method works by systematically modifying feature values and assessing changes in the model's predictions on the altered dataset. While this process is relatively straightforward for tabular data—where multiple features can be modified simultaneously—it poses unique challenges for GNNs. In GNNs, the message passing mechanism updates a node’s embedding using information from its neighbors, which is transmitted via edges and aggregated \cite{Bronstein_Bruna_Cohen_Veličković_2021}. Consequently, a node’s prediction is influenced by its neighbors. If many nodes are modified at once (as is often done in tabular ALE calculations), they may inadvertently influence one another’s predictions, introducing unwanted artifacts. On the other hand, modifying nodes one at a time is computationally expensive and may discourage practitioners from using ALE in GNNs.

This work aims to quantify the extent of the effect described above and answer a key question: does ignoring these inter-node dependencies—by calculating ALE as is done for tabular data—substantially distort the explanation? To answer this, we compute ALE estimates both by ignoring and by accounting for inter-node influences, and we compare the results from the two approaches. Additionally, we evaluate ALE under varying parameter configurations to understand how these settings affect estimation accuracy.

\section*{Methods}
\subsection*{Accumulated Local Effects} Accumulated Local Effects (ALE) plots provide a way to visualize the effect of a feature on the predictions of a machine-learning model by accumulating local changes in the predictions as the feature values vary \cite{Apley_Zhu_2020}. It is an alternative to Partial Dependence Plots \cite{Friedman_2001} and addresses some of its limitations, such as the sensitivity to feature correlations and the inability to accurately capture interactions between features.
The core idea is to measure the local effect of a feature by looking at the changes in predictions when the feature value changes slightly, and then accumulating these changes across the range of the feature. In this expression, the derivative \(f^S(X_S, X_C)\) represents the local effect of \(X_S\) on the model prediction, and this effect is accumulated over the range from \(x_{\text{min}, S}\) to the current value \(x_S\):
\begin{equation}
g_{S, ALE}(x_S) = \int_{x_{\text{min}, S}}^{x_S} \mathbb{E}[f^S(X_S, X_C) | X_S = z_S] dz_S - \text{constant}
\label{eq:ale_formula}
\end{equation}
$X_C$ refers to other features, whose impact is not measured.

The empirical estimation of Accumulated Local Effects is given by:

\begin{equation}
\hat{g}_{S}(x_S) \equiv \sum_{h} \frac{1}{n_S(h)} \sum_{\{i: x_{i,S} \in N_S(h)\}} \left[ f(z_{h,S}, x_C) - f(z_{h-1,S}, x_C) \right]
\label{eq:ale_estimation}
\end{equation}
In this formula, we divide the feature's support into intervals $h$ and sum over them (first summation). The second summation aggregates the local differences in the model's predictions as the feature \(X_S\) transitions from one interval $h$ to another. Values \(z_{h,S}\) and \(z_{h-1,S}\) correspond to the border of the feature \(x_S\) interval, and \(n_S(h)\) represents the number of observations within the \(h\)th interval.

Originally, ALE is centered by subtracting ALE \cite{Apley_Zhu_2020} averaged over all possible values of the feature, making it easier to interpret the contribution of each feature relative to its average effect. For the sake of simplicity of analysis, this procedure will not be applied here.

\subsection*{Modification of ALE for link prediction}
In the task of link prediction, the model returns the probability of an edge existing between two given nodes, $v$ and $u$. This requires a slight adjustment of the ALE method. Instead of modifying features for both nodes involved in the potential link, we focus on altering the features of only one node, which we designate as $v$. The other node, $u$, remains unmodified. In this way, ALE visualizes the effect of the node feature's value on the existence of edges between the modified node and the rest of the dataset.

Graph datasets can be large. Due to this, averaging across all of the nodes present in the dataset could not be feasible. Hence, we take only a subset of size $m$ of nodes, for which we modify the feature $X_S$ we are interested in. We then choose the subset $U$ of size $k$ of nodes, against which we evaluate the link probability for each modified node.

Estimation of ALE from Eq.~(\ref{eq:ale_estimation}) is modified in Eq.~(\ref{eq:ale_estimation_link_pred_k_m}) in order to account for the link prediction task and averaging over only a subset of the nodes:
\begin{equation}
\hat{g}_{S}(x_S) \equiv \sum_{h} \frac{1}{k}\sum_{u \in U} \frac{1}{m} 
\sum_{\substack{v(x_{i,S}, x_{i, C}): \\ x_{i,S} \in N_S(h)}} 
\left[ f(v(z_{h,S}, x_{i, C}), u) - f(v(z_{h-1,S}, x_{i, C}), u) \right]
\label{eq:ale_estimation_link_pred_k_m}
\end{equation}
The sum over $v$ is taken over the nodes with $X_S$ in the interval $h$.  The middle sum (which did not appear in Eq.~(\ref{eq:ale_estimation})) is taken over nodes $u$, which can have an edge with $v$. It is divided by the number of these nodes. 

\begin{figure}[!hb]
    \centering
    \includegraphics[width=.8\textwidth]{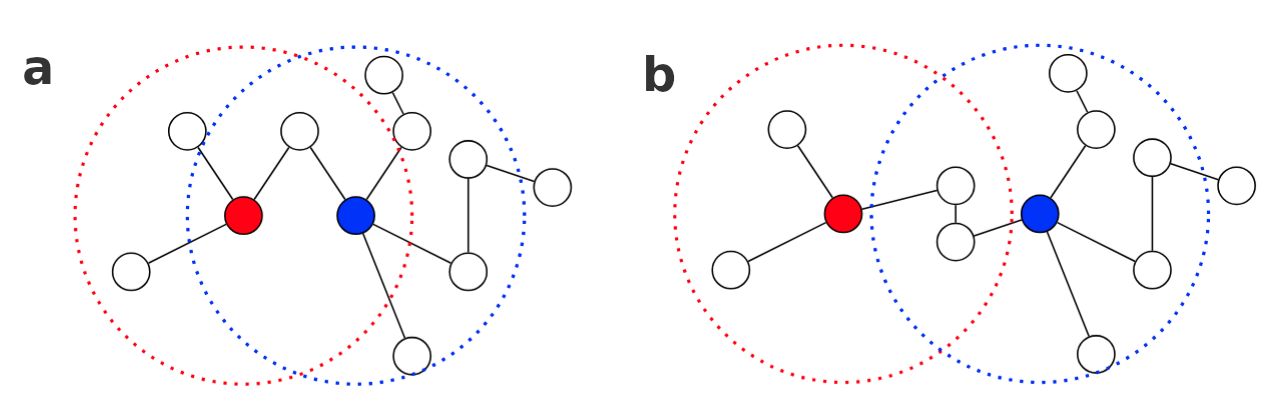}
    \caption{(a) An exemplary graph where two modified features (blue and red) would affect each other during inference through the two-layer GNN. (b) Modification of panel a, where the second node was added on the path between the blue and red nodes. The blue and red nodes would no longer affect each other's embedding.}
    \label{fig:affected}
\end{figure}

However, if multiple nodes were modified at once, there could arise a disturbance in the ALE estimation. Figure \ref{fig:affected}a presents a hypothetical scenario in which modified nodes could affect each others' prediction produced by a two-layer GNN similar to the ones used in this article. Figure \ref{fig:affected}b is a modified version where the nodes would not affect each other embedding. All nodes inside of the dotted circle are connected with the node in the center of this circle by a path no longer than 2. Only information from the nodes inside the dotted circle can affect the embedding of the central node produced by a model with two layers of Graph Convolutional Network (GCN) or Graph Attention Network (GAT). In the first layer, information from the node's neighbors is passed through the edges and aggregated. In the second layer, the same happens, but the neighbors' embedding already contains information about neighbors' neighbors.

The more nodes are modified, the bigger the chance that some of them will be connected by a path short enough to influence each other. Hence, the disturbance coming from the interaction of modified nodes should rise with the number of modified nodes.

Due to the computational constraints, we calculated the explanations for values of parameters $k$ and $m$ being the power of 2 between 16 and 1024. 

\subsection*{Algorithms}
To explore the effect of nodes' interaction during message passing on explanation, we implement two versions of ALE. In the first "approximate" version, the node features are treated as the tabular dataset, and for one interval, the model's prediction is computed simultaneously. The fact that they influence each other while message passing is ignored. This version is further called the approximate version. In the second “exact” version, the value of the explained feature is changed for each examined node at a time, in isolation from the other nodes. The algorithms are presented below:

\begin{algorithm}[H]
\caption{ALE Exact Version}
\begin{algorithmic}[1]
\Require Model $M$, Dataset $\mathcal{D}$, Feature index $f$, Number of bins $N$
\Ensure Accumulated Local Effects (ALE) values
\State Initialize empty list $ALE$
\State Divide feature values into $N$ bins
\For{each bin $b_i$}
    \State Get nodes in bin $b_i$
    \For{each node $n_j$ in $b_i$} \Comment{Additional loop in Exact version}
        \State Set feature $f$ of $n_j$ to lower bin edge
        \State Compute prediction $P_{low}$
        \State Set feature $f$ of $n_j$ to upper bin edge
        \State Compute prediction $P_{high}$
        \State Compute difference $D = P_{high} - P_{low}$
        \State Store $D$
    \EndFor
    \State Compute average difference for bin $b_i$ and update $ALE$
\EndFor
\State \Return $ALE$
\end{algorithmic}
\end{algorithm}

\begin{algorithm}
\caption{ALE Approximate Version}
\begin{algorithmic}[1]
\Require Model $M$, Dataset $\mathcal{D}$, Feature index $f$, Number of bins $N$
\Ensure Accumulated Local Effects (ALE) values
\State Initialize empty list $ALE$
\State Divide feature values into $N$ bins
\For{each bin $b_i$}
    \State Get nodes in bin $b_i$
    \State Set feature $f$ of all nodes in $b_i$ to lower bin edge
    \State Compute prediction $P_{low}$
    \State Set feature $f$ of all nodes in $b_i$ to upper bin edge
    \State Compute prediction $P_{high}$
    \State Compute average difference $D = P_{high} - P_{low}$
    \State Update $ALE$ with $D$
\EndFor
\State \Return $ALE$
\end{algorithmic}
\end{algorithm}

\subsection*{Baseline} \label{goldstandard}
There exists a need for some form of a gold standard to which single explanations could be compared. For the synthetic datasets, the function determining probability of a link's existence is given. However, when using real-life datasets, we do not know this underlying relation. If we want to see how the approximate ALE estimation diverges from the exact ALE curve, we could compare it with the exact explanation calculated for a large enough number of points (so for large $k$ and $m$ parameters). Furthermore, we argue that averaging across multiple ALE curves is equivalent to calculating it for the large number of datapoints at once.

The most accurate ALE estimation was created by averaging the exact explanation for different values of $k$ and $m$. For every value of the parameters, the intervals (and the first sum in Eq.~(\ref{eq:ale_estimation_link_pred_k_m})) remain the same. The latter averages cannot be simply added since the sum of averages is not necessarily the average of sums.  However, if the ALE is multiplied by the $km$ used for the calculation of this estimation, only the sums remain, and the expression becomes additive. In this way, we can sum the predictions obtained during every run of the experiment. We divide it by the number of all predictions in the interval and, in this way, obtain ALE combined from multiple small runs. 

In result, the following formula for aggregate exact explanation is obtained: 
\begin{align}
    \frac{\sum_i k_i m_i g_{S, k_i, m_i}(x_S)}{\sum_i k_i m_i} = \frac{1}{\sum_i k_i m_i} \sum_i k_i m_i\sum_{h} \frac{1}{k_i}\sum_{u \in U} \frac{1}{m_i} 
\sum_{\substack{v(x_{i,S}, x_{i, C}): \\ x_{i,S} \in N_S(h)}} 
\left[ f(v(z_{h,S}, x_{i, C}), u) - f(v(z_{h-1,S}, x_{i, C}), u) \right]
    \label{eq:ale_goldstandard} \\ = \frac{1}{\sum_i k_i m_i} \sum_{h} \sum_{u \in U} 
\sum_{\substack{v(x_{i,S}, x_{i, C}): \\ x_{i,S} \in N_S(h)}} 
\left[ f(v(z_{h,S}, x_{i, C}), u) - f(v(z_{h-1,S}, x_{i, C}), u) \right] = g_{S}(x_S)
\end{align}
In this way, every point contributes to the final estimation with the same weight ($\frac{1}{\sum_i k_i m_i}$), instead of the old weight $\frac{1}{k_i m_i}$.

In this way, multiple ALE profiles can be aggregated into one ALE profile corresponding to the ALE, which would be obtained if predictions from multiple runs were calculated during one ALE estimation.

\section*{Datasets}

 \paragraph{Real-world datasets} We used two real-life datasets. The first one is a citation network of 159,734 Artificial Intelligence research papers from the S2ORC corpus \cite{Lo_Wang_Neumann_Kinney_Weld_2020} with 227,565 citations between them, enriched with author affiliation data from OpenAlex \cite{Priem_Piwowar_Orr_2022,Giziński_Kaczyńska_Ruczyński_Wiśnios_Pieliński_Biecek_Sienkiewicz_2024}. The second dataset is CD1-E\_no2 - a 3D vessel graph of mouse brain vasculature containing 1,664,811 nodes and 2,150,326 edges \cite{Paetzold_McGinnis_Shit_Ezhov_Büschl_Prabhakar_Todorov_Sekuboyina_Kaissis_Ertürk_et_al._2021}. On both datasets, the models are trained for the link prediction task. We explain a given node feature's impact on the existence of the link. In the citation dataset, the explained feature is the fraction of authors affiliated with Big Tech companies, allowing for an investigation into the influence of private sector affiliation on citation patterns. In the vessel graph dataset we explain the z-coordinate of nodes to explore the relationship between the vertical position of the point in the brain and vessel connectivity.

\paragraph{Synthetic dataset} To validate our approach in a controlled setting, we generate a synthetic directed graph with known ground truth relationships between node features and edge formation. The synthetic dataset consists of nodes with 5 features sampled from a standard normal distribution, and a designated signal feature which is uniformly distributed in [-1, 1]. Edges are generated using a sparsity-controlled sampling strategy: for each source node u, we randomly select some of the other nodes as candidate targets and form directed edges with probability $p = (x_u + x_v + 2)/4$, where $x_u$ and $x_v$ are the signal feature values of the source and target nodes, respectively. This construction creates a dataset where edge formation depends on both source and target node features in a known, interpretable way, allowing us to verify that our explanation method correctly identifies the signal feature as the primary driver of connectivity patterns. 

However, it is important to point out that the explanation's ground truth might not be equal to the ground truth understood as the underlying relation. ALE is a model-agnostic explanation method that reveals how the model interprets and uses features, rather than uncovering the true relationships hidden in the data. For example, if the model does not accurately recreate a relationship that is known to be the ground truth, ALE should return the model's learned relation between feature and prediction and not the underlying relation. This distinction is crucial for proper interpretation: ALE explanations tell us what the model has learned, which may include biases, spurious correlations, or simplified approximations of reality. Therefore, when a discrepancy exists between the ALE explanation and domain knowledge, this signals a potential issue with the model itself—either in its architecture, training process, or the data it was trained on—rather than a failure of the explanation method. We present the comparison to the underlying ground truth relation on the synthetic dataset to measure ALE’s and model’s combined ability to extract relationship from the data and then we present the comparison to the aggregated exact explanation on real life datasets, where we compare how approximate method diverges from the exact explanation of the model (ignoring whether the model learns the relationship correctly) .

\section*{Models}
The models were trained for the task of link prediction: given two nodes, the model should return if there exists a link between them. A GNN encoder, either two 256-dimensional layers of Graph Convolutional Network (GCN) \cite{Kipf_Welling_2016} or Graph Attention Network (GAT) \cite{Veličković_Cucurull_Casanova_Romero_Liò_Bengio_2018}, was used to obtain node embeddings on real-life datasets, and the dot product of the embeddings was calculated to predict link probability via a sigmoid function. On the synthetic dataset, similar models with two 64-dimensional layers were used. Binary cross-entropy was used as the loss function, with batch normalization \cite{Ioffe_Szegedy_2015} applied after both layers. The models were implemented using PyTorch Geometric \cite{Fey_Lenssen_2019}. 

The negative sampling of edges was performed, with the number of negative samples equal to the number of positive samples. The Citations dataset models were trained for 15 epochs on a CPU, while the CD1-E\_no2 dataset was trained for 50 epochs on a GPU. The models on the CD1-E\_no2 dataset were trained on a GPU L40 with 24GB. Adam optimizer was used. The learning rate for GCN models was $10^{-6}$ and for GAT models was $10^{-5}$. In order to create train/test datasets, a random link split was performed once for each dataset. A batch size of 1024 was used for both datasets.

\begin{table}[h]
\begin{center}
\begin{tabular}{cccc}

\hline
Model & Dataset & F1 & AUC ROC  \\
\hline
GAT & Citations & $0.6900\pm0.0052$ & $0.6382\pm0.0014$ \\  
GCN & Citations & $0.704\pm0.002$ & $0.7648\pm0.0036$ \\
GAT & CD1-E\_no2 & $0.8589\pm0.0004$ & $0.8939\pm0.0021$ \\  
GCN & CD1-E\_no2 & $0.8526\pm0.0006$ & $0.8836\pm0.0025$ \\
\hline

\end{tabular}
\caption{Metrics for models trained on the Citation and CD1-E\_no2 datasets with the confidence interval calculated across 5 runs.}
\label{table:combined_metrics}
\end{center}
\end{table}

Datasets on the synthetic dataset achieved AUC of $0.7305 \pm 0.0077$ for the GAT model and $0.7297 \pm 0.0076$ for GCN model.

We did not use deeper models, because our initial experiments showed that models with greater number of layers did not exhibit superior performance. Moreover, with greater number of layers appears the risk of oversmoothing.\cite{Rusch_Bronstein_Mishra_2023}, due to which the performance of a GNN often does not increase with a greater number of layers.
However, in the approximate version of ALE, the interaction of modified nodes is more probable, the deeper the network, and there is a possibility that undesirable effects in the approximate version will be greater when the models are deeper. To check whether it is indeed true, we run an experiment with networks with 2, 3, 4 and 5 GCN layers with similar overall number of parameters on the Citations dataset. Five models with respectively 2, 3, 4 and 5 GCN layers were trained on the Citations dataset. Each of the models is later explained 5 times with both methods.

\section*{Results}

\subsection*{Synthetic dataset}
\begin{figure}[!ht]
    \centering
    \includegraphics[width=\textwidth]{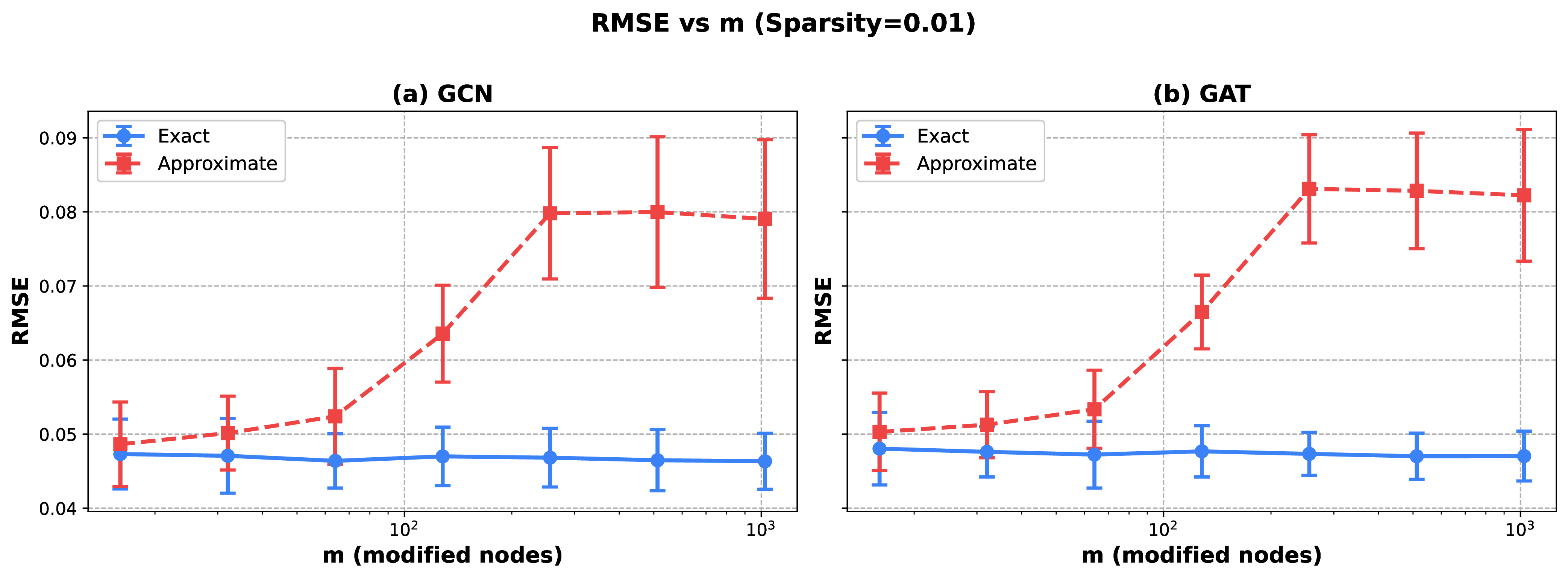}
    \caption{Average Root Mean Square Error between ALE and the underlying relation in the synthetic dataset versus the number of modified nodes $m$ for graph sparsity equal to 0.01 and for (a) GCN model, (b) GAT model. The errorbars correspond to the standard deviation across explanations for different models.}
    \label{fig:rmse_lineplot}
\end{figure}

We measured the divergence from the underlying linear relation in the synthetic dataset using RMSE between the computed explanations and ground truth values. Figure~\ref{fig:rmse_lineplot}a demonstrates that the approximate method's accuracy degrades substantially as the parameter $m$ increases. In contrast, the exact algorithm maintains its fidelity with larger $m$ values (see Fig. \ref{fig:heatmap} in the Supplementary Information). A similar effect was not observed for the $k$ parameter. This divergence is consistent across both GCN and GAT architectures, as shown in the left and right panels, respectively. The effect manifests across all tested graph sparsity levels (0.001, 0.01, 0.1), indicating that the degradation is inherent to the approximation scheme rather than an artifact of graph structure (see Figs. \ref{fig:rmse_lineplot1} and \ref{fig:rmse_lineplot2} in the Supplementary Information). The standard deviation between different models is increasing with the $m$ parameter for the approximate, but not for the exact method.

\subsection*{$\chi^2$ Test} To determine whether the ALE curves from both methods differ or if they can be used interchangeably, we applied a $\chi^2$ test adjusted for comparing arbitrary curves \cite{Hristova_Wimley_2023}. The null hypothesis assumed that the ALE profiles came from the same distribution. As recommended by Hristova and Wimley\cite{Hristova_Wimley_2023}, the degrees of freedom were set to the number of points in the curve. At a significance level of $\alpha=0.05$, the null hypothesis would be rejected if the $\chi^2$ value exceeded 11.07. For the Citations dataset, the $\chi^2$ values for the ALE curves were 7.165 for the GCN model and 5.413 for the GAT model. For the CD1-E\_no2 dataset, the $\chi^2$ values were 17.439 for the GCN model and 1.296 for the GAT model. A statistically significant difference between the curves was observed only for the GCN model trained on the CD1-E\_no2 dataset.

\begin{figure}[!ht]
    \centering
    \includegraphics[width=\textwidth]{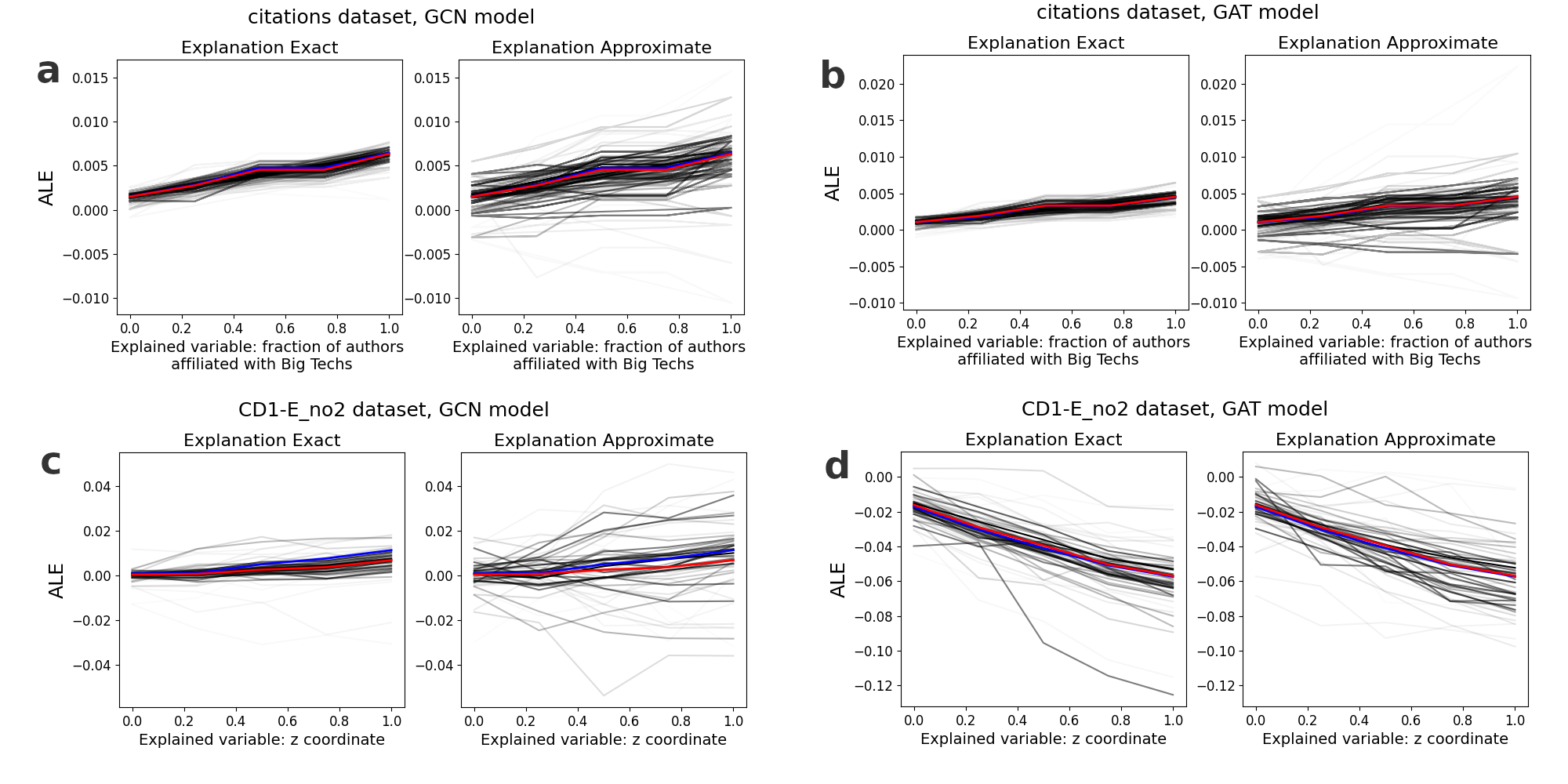}
    \caption{ALE curves calculated for the fraction of authors affiliated with Big Techs for CGN model (a), GAT model (b) and for the $z$ coordinate in CD1-E\_no2 for the CGN model (c), GAT model (d). The hue corresponds to the number of edges taken into account during calculating ALE profile. The red line is the average of the exact predictions weighted with the number of predictions which corresponds to the exact explanation calculated with a large number of nodes, and the blue line is the weighted average of approximate predictions.}
    \label{fig:ale}
\end{figure}

\subsection*{Permutation Test} We conducted a permutation test to assess whether the exact and approximate explanations differ significantly. The null hypothesis stated that both groups were sampled from the same distribution. The test statistic was the weighted average ALE profile, and the difference was measured as the root mean squared error between the averaged profiles of the two groups. The p-value was the percentage of tests where the difference between the test statistics of the two groups exceeded that of the original group split. A total of $n=10,000$ splits of ALE curves into two groups were randomly generated.

For the Citations dataset, the p-value for explanations of GCN model was 0.407, and for GAT model - 0.898. For the CD1-E\_no2 dataset, it was 0.195 for GCN model and 0.155 for GAT model. No p-value was smaller than the significance level $\alpha=0.05$. Hence, the null hypothesis stating that samples are taken from the same distribution was not rejected in any case.

It should be noted that the lack of a significant difference does not imply that both curves come from the same distribution. As can be seen in Fig. \ref{fig:ale}, approximate explanations seem to have greater variance. However, if the difference between two distributions is small enough, using an approximate method can be justified.

\begin{figure}
    \centering
    \includegraphics[width=\linewidth]{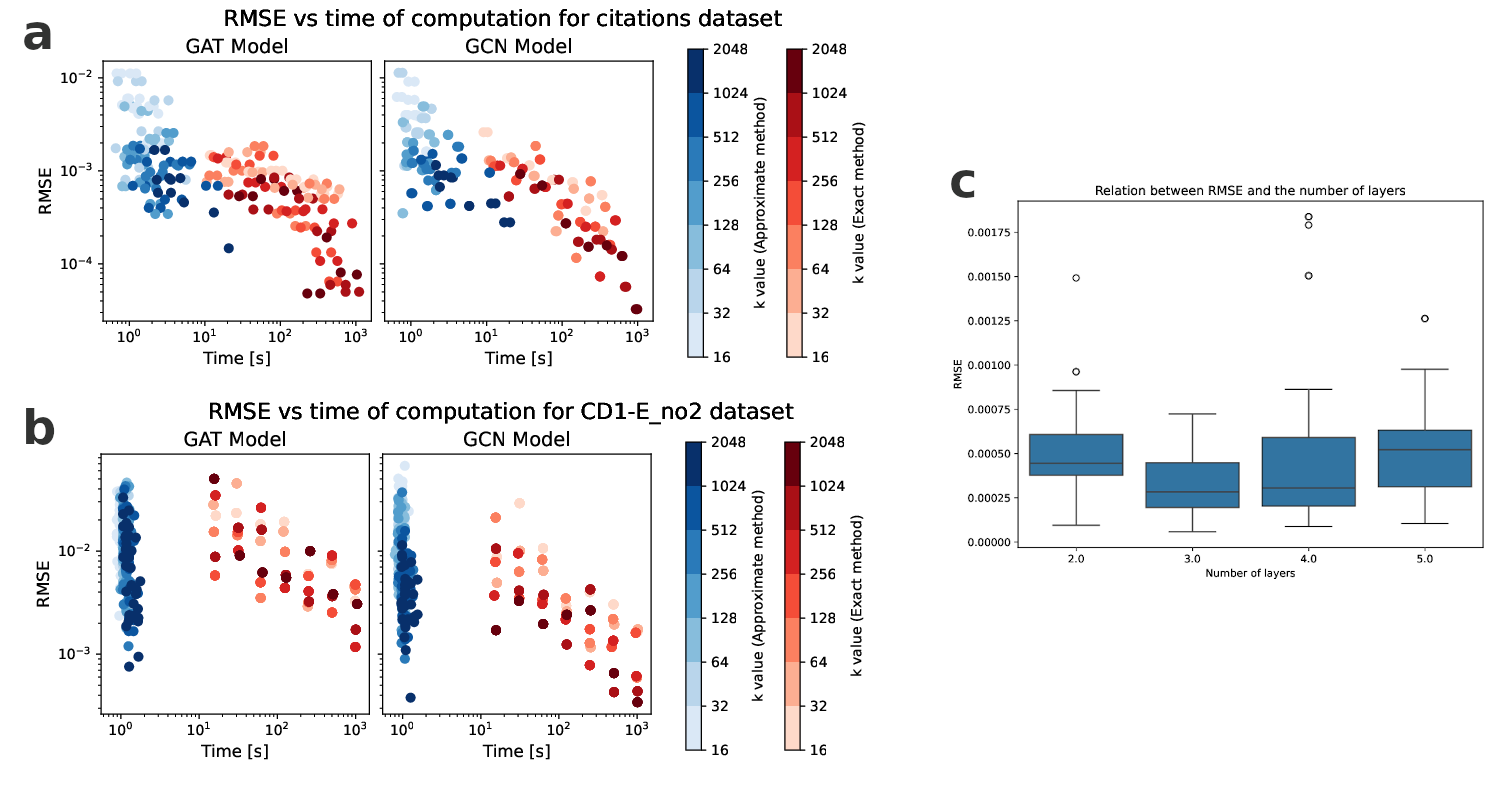}
    \caption{RMSE between runs and aggregate exact explanation plotted against time of explanation for Citations dataset (a) and CD1-E\_no2 dataset (b). The red dots are the exact ALE and the blue dots are the approximate ALE. The hue corresponds to $k$. The time of exact explanation is roughly proportional to the $m$ parameter. (c) The RMSE between the approximate and exact explanation in relation to the number of layers of the GCN models trained on the Citations dataset}
    \label{fig:rmse}
\end{figure}

There exists a bigger variability for the single runs of approximate ALE than for the exact ALE. It can be observed in Fig. \ref{fig:ale} -- the smaller the number of nodes taken into account, the stronger this effect. Figures \ref{fig:rmse}a and \ref{fig:rmse}b show that the higher the $k$ parameter, the smaller the RMSE between the approximate ALE and the aggregate exact explanations. However, this effect is not visible for the exact ALE. From the same plots, we conclude that the higher the time of explanation (proportional to the $m$ parameter), the better the exact ALE. 

\subsection*{Network's depth impact}
The RMSE between approximate explanations and aggregate exact explanation can be found in Fig. 
\ref{fig:rmse}c. Five models with respectively 2, 3, 4 and 5 GCN layers were trained on the Citations dataset. Each of them was explained 5 times. We do not observe a clear dependence between the number of layers and the error of the explanations.

\section*{Discussion}
The results on the synthetic dataset show that increasing the $m$ parameter increases the error between the explanation and the underlying relation encoded in the data for the approximate, but not for the exact method. This confirms the existence of the disturbance stemming from modifying many nodes at once. 

The results of $\chi^2$ tests showed that in 3 out of 4 different models, the results obtained with both methods were not significantly different.
The permutation test did not show differences between the approximate and exact methods' results in any case. 

For the exact method of ALE calculation in link prediction tasks in GNNs, it is more beneficial to increase the $m$ parameter than $k$ parameter. In this way, more nodes with modified feature's value are taken into account. This comes at the time expense since computation scales linearly with the number of nodes in the interval. Although the approximate method of explanation has greater variability between single runs, it could be used in time-sensitive scenarios. This variability can be reduced by increasing the predicted number of edges (by increasing $k$ and $m$ parameters).

\subsection*{Relationship between explanations and real-life phenomena}

Figures \ref{fig:ale}a,b show that the probability of being cited increases with the fraction of authors affiliated with Big Techs. This relation between the article's popularity and authors' affiliation is consistent with the literature on this topic \cite{Färber_Tampakis_2024}. However, it does not support the conclusions of the PageRank and node degree analysis of Giziński et al. \cite{Giziński_Kaczyńska_Ruczyński_Wiśnios_Pieliński_Biecek_Sienkiewicz_2024}, which revealed that the most popular were articles with authors affiliated both with Big Techs and Academia. The latter analysis was performed on the Citations dataset, which is also used in this work.
This discrepancy between the two analyses could stem from differences in the methodologies used or from the possibility that our models did not capture more complex relationships present in the data. Additionally, Fig. \ref{fig:ale}c shows that the probability of a link forming increases with the z-coordinate, whereas Fig. \ref{fig:ale}d suggests the opposite — a decreasing probability. This contradiction may arise from the two models learning opposing relationships between the z-coordinate and the likelihood of an edge forming between nodes.

These results underscore the utility of ALE (Accumulated Local Effects) in assessing how node features influence GNN predictions. However, it's important to note that explainability tools like ALE reveal only what the models have learned, not necessarily the underlying real-world phenomena.
    
\section*{Conclusions}
This article provides an analysis of how ALE can be adjusted to work for link prediction tasks on GNN models. It proposes an approximation to speed up the process and researches the impact of this approximation on the accuracy of the explanation.
We argue that ALE methods uniquely complement other GNN-explaining methods by providing a way to visualize node feature's impact on the prediction instead of the graph's elements which were important for the prediction or which alterations could change the prediction. Moreover, we believe that using a method applied outside of GNNs might be more intuitive and easier for people who did not have contact with GNN explanations but are accustomed to explanation methods outside of GNNs.

We show how, in most cases, the explanations produced with the approximate method are not significantly different from the explanations produced with the exact method. This leads us to the conclusion that, especially in time-sensitive situations, the node interaction effects can be ignored. 
The approximate explanations vary more than the exact ones, but averaging many approximate explanations limits this effect, while still providing improvement in time. 
The $k$ parameter - the number of nodes possibly having an edge with a modified node - has a clear impact on the accuracy while using the approximate method, but we do not observe a similar impact with the exact method. The network depth did not have a clearly visible impact on the approximate explanation's accuracy. 

In the future, a similar analysis could be made for different tasks like node classification, edge classification, or graph classification. Additionally, exploring how ALE could be adapted for dynamic and temporal graph learning tasks \cite{LiDonguanTan2023} would be valuable, as these settings introduce additional complexity through time-evolving graph structures. Similar disturbance in ALE calculation stemming from modifying multiple points at the same time could also appear in other architectures, for example, in the transformer \cite{Vaswani_Shazeer_Parmar_Uszkoreit_Jones_Gomez_Kaiser_Polosukhin_2023}.

ALE applies to continuous variables, limiting its applicability to categorical node features. This restricts the analysis to a smaller number of node features in graph datasets. However, its adjustments to categorical features or similar methods like PDP will also be prone to the disturbance described in this paper.

\section*{Supplementary Information}

\begin{figure}[!ht]
\begin{tabular}{c|c}
     \includegraphics[width=.5\textwidth]{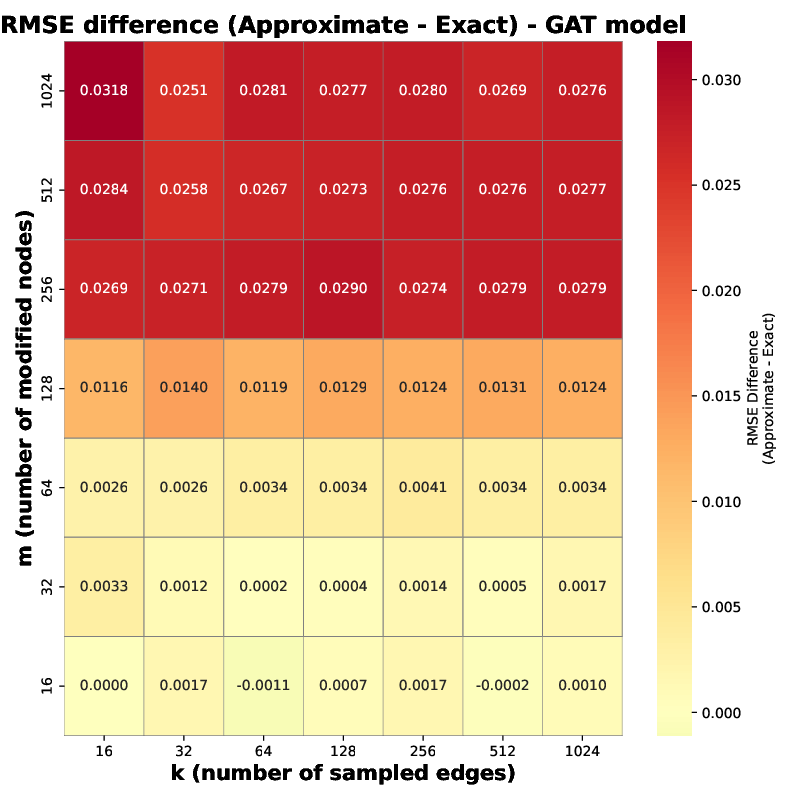} &  
     \includegraphics[width=.5\textwidth]{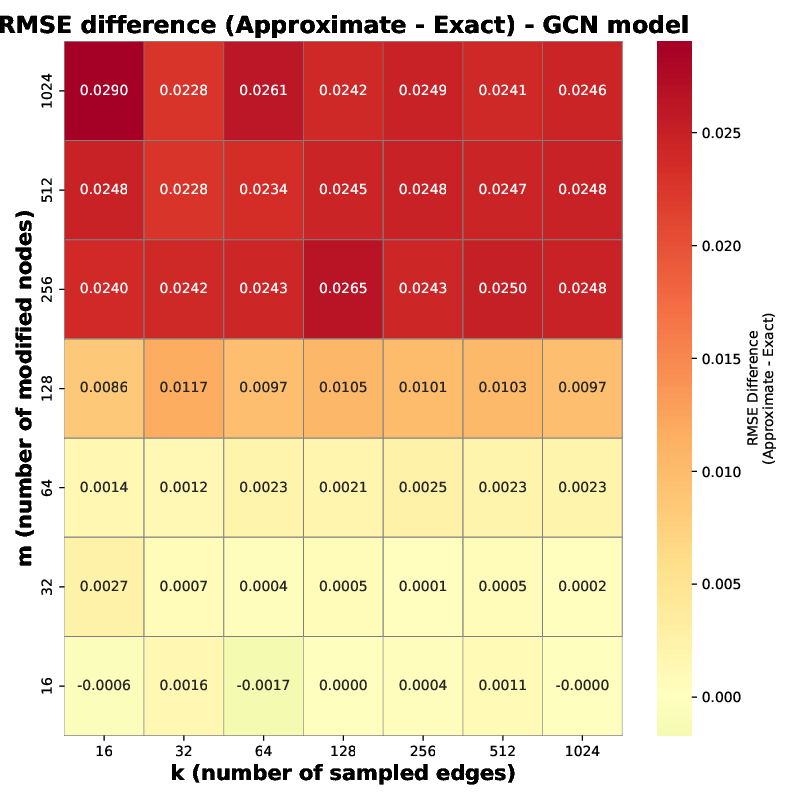}\\
\end{tabular}
    \caption{Heatmaps of Root Mean Square Error difference between the approximate and exact method for GAT model (left) and GCN model (right) versus the number of sampled edges $k$ (horizontal axis) and the number of modified nodes $m$ (vertical axis).}
    \label{fig:heatmap}
\end{figure}

\begin{figure}[!ht]
    \centering
    \includegraphics[width=\textwidth]{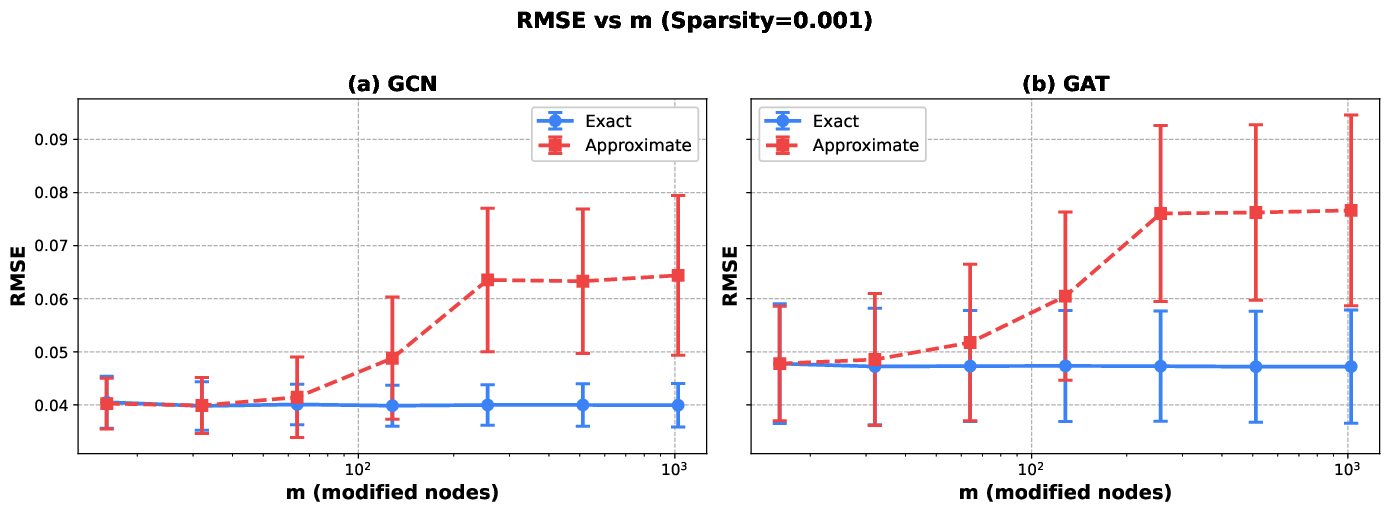}
    \caption{Average Root Mean Square Error between ALE and the underlying relation in the synthetic dataset versus the number of modified nodes $m$ for graph sparsity equal to 0.001 and for (a) GCN model, (b) GAT model. The errorbars correspond to the standard deviation across explanations for different models.}
    \label{fig:rmse_lineplot1}
\end{figure}

\begin{figure}[!ht]
    \centering
    \includegraphics[width=\textwidth]{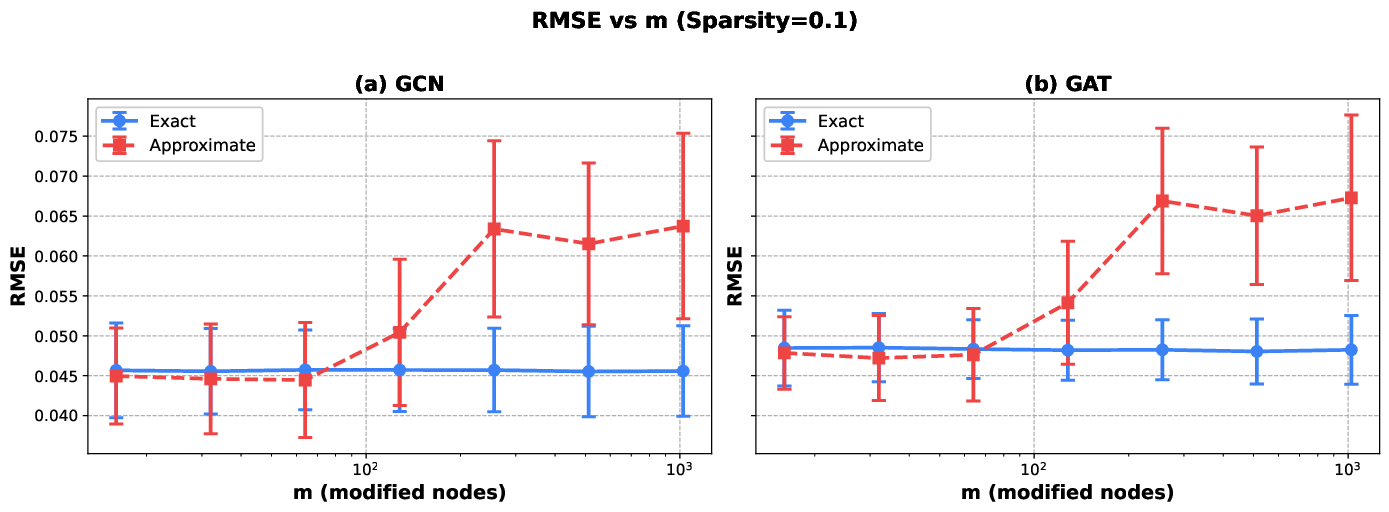}
    \caption{Average Root Mean Square Error between ALE and the underlying relation in the synthetic dataset versus the number of modified nodes $m$ for graph sparsity equal to 0.1 and for (a) GCN model, (b) GAT model. The errorbars correspond to the standard deviation across explanations for different models.}
    \label{fig:rmse_lineplot2}
\end{figure}

\bibliography{sn-bibliography}

\section*{Code availability}
The code used to produce the results presented in this work is available at: \href{https://github.com/Kaczyniec/ALE-and-GNNs}{https://github.com/Kaczyniec/ALE-and-GNNs}.

\section*{Acknowledgements}
\textbf{P.K.} and \textbf{J.S.} acknowledge support by POB Cybersecurity and Data Science of Warsaw University of Technology within the Excellence Initiative: Research University (IDUB) programme. Addionally \textbf{J.S.} acknowledges support by the European Union under the Horizon Europe grant
OMINO – Overcoming Multilevel INformation Overload (grant number 101086321, \href{https://ominoproject.eu}{https://ominoproject.eu}). Views and opinions expressed are those of the authors alone and do not necessarily reflect those of the European Union or the European
Research Executive Agency. Neither the European Union nor the European Research Executive Agency can be held responsible
for them.

\section*{Author contributions}
\textbf{P.K.}, \textbf{J.S.}, and \textbf{D.S.} conceived and planned the study. \textbf{P.K.} wrote the draft version of the manuscript, performed experiments, and wrote the code. All authors reviewed the manuscript.

\section*{Competing interests}
The authors declare no competing interests.

\end{document}